\documentclass[twocolumn,prd,showpacs,amsmath]{revtex4}
\usepackage{epsfig,amsfonts,amsbsy}
\def\nn{\nonumber}
\def\lb{\label}
\def\ci{\cite}
\renewcommand{\Ref}[1]{(\ref{#1})}

\newcommand{\p}{\partial}
\newcommand{\N}{\nabla}

\def\a{\alpha}
\def\b{\beta}
\def\g{\gamma}
\def\d{\delta}
\def\D{\Delta}

\def\e{\epsilon}

\def\th{\theta}
\def\s{\sigma}

\def\bra{\langle}
\def\ket{\rangle}
\def\l{\left}
\def\r{\right}
\def\f{\frac}
\def\MO{\mathcal {O}}
\def\MR{\mathcal {R}}

\def\cw{{\widetilde c}_W}
\def\aw{{\widetilde a}_W}
\def\bw{{\widetilde b}_W}
\begin{document}
\title{Interior Metric of Slowly Formed Black Holes in a Heat Bath} 
\author{Hikaru Kawai}
\affiliation {
Department of Physics and Center for Theoretical Physics, National Taiwan University, Taipei 106, Taiwan\\
Physics Division, National Center for Theoretical Sciences, Taipei 10617, Taiwan
}
\author{Yuki Yokokura}
\affiliation {iTHEMS Program, RIKEN, Wako, Saitama 351-0198, Japan}

\begin{abstract}
We study a spherical black hole formed slowly in a heat bath in the context of ordinary field theory, 
which we expect to have the typical properties of black holes. 
We assume that the matter field is conformal and that the metric satisfies the semi-classical Einstein equation
$G_{\mu\nu}=8\pi G \langle \psi| T_{\mu\nu}|\psi \rangle$, where $|\psi \rangle$ is the wave function of the matter field.
Then, as a necessary condition, its trace part must be satisfied, 
$G^\mu{}_\mu=8\pi G \langle \psi| T^\mu{}_\mu|\psi \rangle$, 
whose right-hand side is independent of $|\psi \rangle$ and is determined only by the metric through the 4-dimensional Weyl anomaly.
With some physically reasonable assumptions, this equation restricts the interior metric to a certain class.
Such metrics are approximately warped products of $AdS_2$ and $S^2$ with almost Planckian curvature. 
Among them, we find one that is consistent with Hawking radiation 
and is smoothly connected to the exterior Schwarzschild metric slightly outside the Schwarzschild radius. 
This leads to a picture that the black hole is a dense object with a surface (not a horizon),
which evaporates due to Hawking-like radiation when taken out of the bath. 
Other solutions represent objects that are larger in size than a black hole, which may be useful for analyzing ultra-dense stars.
\end{abstract}

\maketitle
\section{Introduction}\lb{s:Intro}
What is a black hole?  
The answer is still unclear 
because observationally nothing is known about the interior \ci{LIGO,EHT,Cardoso} 
and theoretically the information problem \cite{Hawking2} caused by the evaporation of black holes \ci{Hawking} has not been resolved. 
Therefore, at this stage, the most conservative definition of a black hole is that it is an object formed 
by gravitational collapse that reduces matter to a minimum size that does not violate quantum 
theory \ci{Gerlach, Frolov, tHooft,Vachaspati, Barcelo1,KMY,Mers1,KY1,KY2,Barcelo2,Ho3,Terno1,KY3, Malafarina,KY4,Minic}.

In general, 
a black hole with mass $M=\f{a}{2G}$ can be in equilibrium with the heat bath 
of temperature $T_H=\f{\hbar}{4\pi a}$ \ci{GH}, 
and when taken out of the bath it evaporates in the time scale $\Delta t=\MO(\f{a^3}{l_p^2})$, 
where $a\gg l_p\equiv \sqrt{\hbar G}$. 
In this paper, we use this property 
and consider a spherically-symmetric black hole which has been formed slowly in a heat bath. 
That is, we grow a small black hole into a large one by varying the temperature and (finite) 
size of the bath in a sufficiently long time scale $\D t$ \ci{KY1}. 
Here, we assume that $\D t$ is larger than $\MO(\f{a^3}{l_p^2})$,
but not so large that it exceeds all powers of $a$: 
$\MO(\f{a^3}{l_p^2})\ll \D t < \MO(\f{a^{k+1}}{l_p^k})$ for some $k>2$. 
The meaning of this time scale will be explained below \Ref{freez_t}. 
We can regard such a black hole as a typical one, 
and a more general black hole may be understood as a perturbation from it. 

Let us discuss what happens to the fields during the slow formation. 
First, radiation comes slowly from the heat bath, 
and the spacetime curves gradually. 
Then, the temporal change of the metric causes particle creation and energy release \ci{BD,Pre,KMY}.  
At the same time, quantum fluctuations of various modes 
in the vicinity of the black hole are induced by the curved geometry, resulting in
negative energy flow \ci{DFU,Fulling}, pressures in various directions \ci{Howard,KY4}, and so on.  
These effects occur at each stage of such a formation process.

One way to incorporate such complex effects is to solve 
the semi-classical Einstein equation 
\begin{equation}\lb{Einstein}
G_{\mu\nu}=8\pi G \bra \psi |T_{\mu\nu}|\psi \ket
\end{equation}
in a self-consistent manner \ci{KMY,KY1,KY2,Ho3,KY3,KY4}. 
In this paper, 
we consider conformal matter fields 
and focus on the trace part 
using the 4-dimensional (4D) Weyl anomaly \ci{BD,Duff}: 
\begin{equation}\lb{trace}
G^\mu{}_{\mu}=8\pi G \bra \psi |T^\mu{}_{\mu}|\psi \ket=8\pi G \hbar \l(c_W {\cal F}-a_W{\cal G}+b_W \Box R \r).  
\end{equation}
Here, we have 
${\mathcal F}\equiv C_{\a\b\g\d}C^{\a\b\g\d}=R_{\a\b\g\d}R^{\a\b\g\d}-2R_{\a\b}R^{\a\b}+\f{1}{3}R^2$ 
and 
${\mathcal G}\equiv R_{\a\b\g\d}R^{\a\b\g\d}-4R_{\a\b}R^{\a\b}+R^2$, 
and $c_W$ and $a_W$ are determined by the matter content 
of the theory for small coupling constants, 
while $b_W$ depends on the coefficients of 
the $R^2$ and $R^{\mu\nu}R_{\mu\nu}$ terms of the gravity action. 
There are two important points with \Ref{trace}. 
One is that \Ref{trace} holds independently of $|\psi\ket$ 
because the anomaly is an operator identity, so any solution of \Ref{Einstein} must satisfy it. 
The other is that the 4D Weyl anomaly comes from quantum fluctuations of all modes with arbitrary angular momentum, 
and the metric that exactly satisfies \Ref{trace} contains the various effects above.

Before starting the detailed analysis of \Ref{trace}, 
let us estimate the magnitude of the curvature $\MR$ of the black hole with mass $M=\f{a}{2G}$. 
From \Ref{trace}, we have 
\begin{equation}\lb{Req}
\MR \sim l_p^2 N \MR^2, 
\end{equation}
where we have assumed that $G^\mu{}_\mu\sim \MR$ and $ {\cal F},{\cal G}, \Box R\sim \MR^2$ 
and $N$ stands for the degrees of freedom of the matter fields (i.e. $c_W,a_W,b_W$). 
This has two types of solutions. 
One type is approximated as $\MR\sim 0$, more precisely, 
\begin{equation}\lb{Rsmall}
\MR\sim \f{Nl_p^2}{a^4}, 
\end{equation}
which is very small compared to $\f{1}{l_p^2}$. 
Here, we have used the Schwarzschild metric \Ref{Sch} and the anomaly formula (right-hand side of \Ref{trace}): 
$\bra T^\mu{}_\mu\ket =\f{12\hbar(c_W-a_W)a^2}{r^6} \sim \f{\hbar N}{a^4}$ for $r\sim a$. 
This should be represented by a slightly modified Schwarzschild metric  
including the small backreaction of dilute radiation and vacuum polarization around $r=a$.
The other one balances the both sides of \Ref{Req} exactly, 
\begin{equation}\lb{R}
\MR\sim \f{1}{Nl_p^2},
\end{equation}
which is non-perturbative for $\hbar$. 
The energy scale is almost Planckian, $\MO\l(\f{m_p}{\sqrt{N}}\r)$, 
where $m_p\equiv \sqrt{\f{\hbar}{G}}$. 
However, if $N$ is fairly large, it is still less than the Planck scale. 
Therefore, the semi-classical Einstein equation \Ref{Einstein} is valid, albeit barely 
(except when the black hole is very small such as $a\sim l_p$).

In general, the one-loop effective action consists of nonlocal and local terms, 
but the former is indefinite in that any of the latter may be added. 
In the case of $N$ matter fields, the purely nonlocal part of the former is proportional to $N$. 
The latter consists of the Einstein-Hilbert term $R$ and higher-derivative terms such as $R^2$ and $R^3$, 
whose coefficients can be chosen arbitrarily by finite renormalization.  
In this paper, we choose (at a renormalization point $\sim \f{m_p}{\sqrt{N}}$) the renormalized Newton constant $G$ to be $\MO(N^0)$, 
and take the renormalized coupling constants of the higher-derivative terms to consist of two parts, $\MO(N^0)$ and $\MO(N^1)$. 
Furthermore, we interpret $\bra T_{\mu\nu}\ket$ in \Ref{Einstein} 
as being given by the local terms with coefficients of $\MO(N^1)$ and the purely nonlocal term. 
Then, $\bra T_{\mu\nu}\ket$ is proportional to $N$ in total. 
On the other hand, 
all the higher-derivative corrections to the left-hand side of \Ref{Einstein} have coefficients of $\MO(N^0)$. 
Therefore, when \Ref{R} is satisfied, 
the left-hand side with such corrections becomes 
$\MO(N^0)R+\MO(N^0)R^2+\cdots \sim \MO(N^{-1})+\MO(N^{-2})+\cdots$, 
and for a large $N$, we only need to consider the Einstein-Hilbert term $R$, 
which justifies the use of of \Ref{Einstein}. 
(In this case, the higher-derivative corrections to $\bra T^\mu{}_\mu\ket$ 
in the right-hand side of \Ref{trace} can also be ignored \ci{foot:no_correct}.)

In this paper, we will restrict our discussion to such a special model \ci{KY4}. 
Then, we can safely ignore the corrections due to the higher-derivative terms 
and obtain a model of black holes that incorporates the backreaction 
of all modes with arbitrary angular momentum in a non-perturbative manner for $\hbar$.

\begin{figure}[h]
\begin{center}
\includegraphics*[scale=0.12]{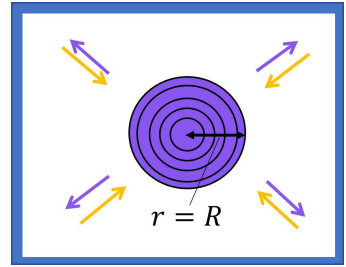}
\caption{The slowly-formed black hole as a dense object formed in a finite-size heat bath.
It has a surface as the boundary at $r=R(>a)$, instead of a horizon.}
\label{f:BH}
\end{center}
\end{figure}

Then, what does \Ref{R} mean physically?
A spherically symmetric solution satisfying \Ref{R} represents 
a spherical dense object because $\bra T^\mu{}_\mu\ket\sim \f{{\cal R}}{G} \sim \f{\hbar}{Nl_p^4}$ means 
that some components of $\bra T^\mu{}_\nu\ket$ are much larger than the order of \Ref{Rsmall}. 
Therefore, it is a good candidate for the interior metric of a black hole that has been formed slowly in the heat bath. 
In fact, at the stage when the mass reaches $\f{a'}{2G}$ in the formation process, 
the quantum fluctuations of the various modes near $r=a'$ condense to form a layer like a ``brickwall" \ci{brick}. 
This happens at each stage, 
and concentric brickwall-like layers continuously pile up at each $r$, 
forming a dense object with mass $\f{a}{2G}$ and curvature \Ref{R}, as in Fig.\ref{f:BH}. 
We call such an object a \textit{slowly-formed black hole}. 

Thus, \Ref{trace} can describe both the dilute exterior region with \Ref{Rsmall} 
and the dense interior region with \Ref{R}. As will be shown later, 
the two regions are smoothly connected at a radius, $r=R$, 
where $R \approx a$ and $R>a$. 
Therefore, instead of a horizon, this object has a \textit{surface} at $r=R$ 
as the boundary between the two regions.

This picture has already been obtained by several self-consistent analyses of \Ref{Einstein}, 
considering the time evolution of a collapsing matter 
and including the backreaction from the particle creation during the collapse \ci{KMY,KY1,KY2,KY3,KY4}.
The present paper provides another evidence to the picture by a simpler approach. 

The strategy is to solve directly \Ref{trace} for $r\gg l_p$, 
whose results hold for \textit{any} state $|\psi\ket$.
In Sec.\ref{s:metric}, we first set an ansatz for the metric 
and study the asymptotic behavior of $\bra T_{\mu\nu}(r)\ket$ for $r\gg l_p$. 
Using it, in Sec.\ref{s:solve}, we solve \Ref{trace} and obtain a class of solutions that satisfy \Ref{R}. 
They are approximately a warped product of $AdS_2$ and $S^2$ 
and have an almost Planckian tangential pressure. 
In Sec.\ref{s:n=2}, 
we see that among the class, there is a special one that is consistent with Hawking radiation.
This describes the interior of the slowly-formed black hole and realizes the above picture. 
The fact that the interior metric holds for any state $|\psi\ket$ 
makes the picture more robust. 
The other solutions in the class represent 
the interior of objects with a size larger than black holes, which may be candidates of ultra-dense stars. 
In Sec.\ref{s:Einstein}, 
we discuss briefly how the interior metric of the black hole satisfies the other components of \Ref{Einstein} 
and explain the origin of the large pressure.
We also study time evolution of quantum matter fields in the obtained background geometry 
and prove that Hawking-like radiation occurs self-consistently. 
In Sec.\ref{s:Con}, we conclude this paper with conclusions and future directions. 

Here, we note that the analysis in this paper is always applicable 
to cases where matter fields are considered to be approximately conformal 
in an almost Planckian energy region.

\section{Study of $g_{\mu\nu}$ and $\bra T_{\mu\nu}\ket$}\lb{s:metric}
\subsection{Mass-independence of the interior metric $g_{\mu\nu}^{(in)}$}
We start by setting the metric of the object with mass $M=\f{a}{2G}$. 
We first consider the dense interior region $(r<R)$. 
Because it is spherically-symmetric and static, 
the interior metric for $r<R$, $g_{\mu\nu}^{(in)}$, can be expressed as 
\begin{equation}\lb{AB}
ds^2=g_{\mu\nu}^{(in)}(x)dx^\mu dx^\nu=-\f{e^{A(r)}}{B(r)}dt^2 + B(r)dr^2 +r^2 d\Omega^2. 
\end{equation}
Generically, $A(r)$ and $B(r)$ are determined by the mass $\f{a}{2G}$ and $l_p$, 
since we are considering conformal matter fields. 
However, we can assume that they are independent of $a$. 
This is motivated by the following two observations.

\textit{(i) Multi-shell model.}--- 
The first one is based on the result of the self-consistent analysis of a model 
that describes the time evolution of a spherical collapsing matter 
including the backreaction of the particle creation during the collapse \ci{KMY,KY3,KY4}. 
Although the main argument of this paper does not depend on this model, 
we can use it to see explicitly that the interior metric does not depend on the mass. 
Here, we give its physical picture (see Appendix \ref{A:model} for details). 

\begin{figure}[h]
\begin{center}
\includegraphics*[scale=0.08]{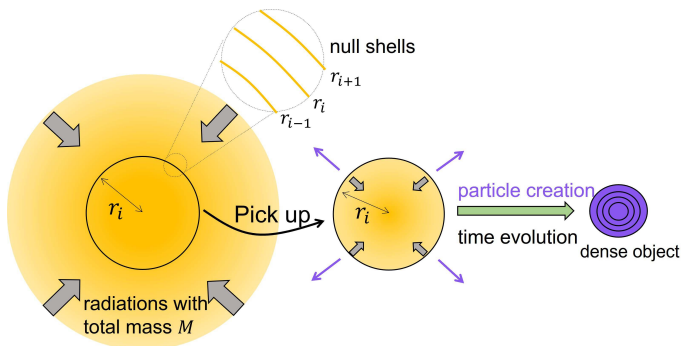}
\caption{A model of spherical radiations as concentric null thin shells. 
$r_i$ denotes the position of the $i$-th shell.}
\label{f:shells}
\end{center}
\end{figure}

Suppose that, during the formation process in the heat bath, 
a lot of radiation (with total energy $M$) comes together in a spherically-symmetric way. 
See Fig.\ref{f:shells}. 
We model this by many concentric null thin shells. 
The shells contract, and the spacetime changes with time, creating particles. 
It can be seen for the case of continuously-distributed shells that 
if we consider the backreaction from the particle creation in a self-consistent manner, 
the shells inside the $i$-th one become a dense spherical object 
and the metric inside approaches a stationary one. Here, the shells outside 
the $i$-th one cannot catch up with it since the shells are null, 
and the metric inside is not affected by the spacetime outside the $i$-th shell because of the spherical symmetry. 
Therefore, the stationary metric cannot depend on the total mass $M$. 
This result can be applied to any shell, and thus the interior metric is not dependent on $M$. 
In fact, it is a special case of the metric to be obtained later. 

\begin{figure}[h]
\begin{center}
\includegraphics*[scale=0.09]{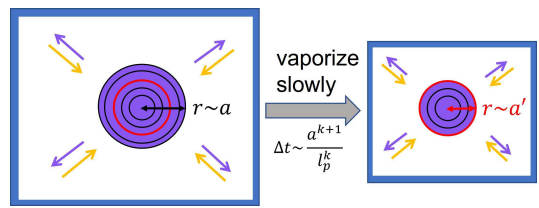}
\caption{Slow change in the heat bath. 
A slowly-formed black hole with size $a$ is vaporized to $a'$ 
in the time scale  $\Delta t \sim \f{a^{k+1}}{l_p^k}$ with $k>2$.}
\label{f:change}
\end{center}
\end{figure}

\textit{(ii) Phenomenological discussion.}---
For the second observation, 
we will use a generic argument and show that $g^{(in)}_{\mu\nu}$ is nearly independent of $a$ \ci{KY1}. 
First, it is natural to assume that 
the relaxation time of a black hole with size $a$ is at most $\D t \sim \f{a^3}{l_p^2}$, 
since the time scale of the evaporation is $\D t \sim \f{a^3}{l_p^2}$. 
Then, a slow change in $\Delta t \sim \f{a^{k+1}}{l_p^k}$ with $k>2$ corresponds to an ``adiabatic" change. 
Next, due to the extremely high density of the slowly-formed black hole (as in Fig.\ref{f:BH}), 
the internal redshift should be very large. 
As we will see later, indeed, it is exponentially large so that the time inside is almost frozen. 
Now, suppose that we control the temperature and size of the heat bath and 
vaporize a slowly-formed black hole with $a$ to $a'(<a)$ in $\Delta t \sim \f{a^{k+1}}{l_p^k}$ with $k>2$. 
See Fig.\ref{f:change}. 
Then, the structure inside $r=a'$ dose not change almost 
because the inside time corresponding to the outside time of order $\f{a^{k+1}}{l_p^k}$
is very small due to the exponentially large redshift. 
This means that $g^{(in)}_{\mu\nu}$ of the black hole with $a'$ 
is almost the same as that of the black hole with $a$. 
Thus, $g^{(in)}_{\mu\nu}$ is almost independent of the total mass. 

Motivated by (i) and (ii), 
we thus assume as a first trial that the interior metric $g^{(in)}_{\mu\nu}$ is $a$-independent \ci{foot:a-dep}. 
As we will see below, 
we can construct a metric that satisfies \Ref{Einstein} and is consistent with this assumption. 

On the other hand, for the exterior dilute region $(R<r)$, 
we ignore the small quantum effects around $r=a$ as a first approximation 
and assume the Schwarzschild metric:
\begin{equation}\lb{Sch}
ds^2=-\l(1-\f{a}{r}\r)dt^2 +\l(1-\f{a}{r}\r)^{-1}dr^2+r^2d\Omega^2.
\end{equation}

\subsection{Asymptotic behavior of $\bra T_{\mu\nu}(r)\ket$}
Next, we examine the asymptotic behavior of $\bra T^\mu{}_{\nu}(r)\ket$ 
when $r$ is large compared with $l_p$ but still inside the object ($l_p \ll r<R$).
We begin with the fact that the mass of the system, $M=\f{a}{2G}$, can be expressed as  
\begin{equation}\lb{mass}
M=4\pi \int^{R}_0 dr r^2 (-\bra T^t{}_t (r)\ket),
\end{equation}
as long as $R\gg l_p$ and $(r^2 \bra T^t{}_t (r) \ket)|_{r\to 0}$ is finite \ci{Landau_C}. 
Note that the $\bra T^\mu{}_{\nu}(r)\ket$ is common to all the slowly-formed black holes
because $g_{\mu\nu}^{(in)}$ does not depend on $a$. 
Then, in order for $M$ to be an increasing function of $R$, 
the energy density $-\bra T^t{}_t (r)\ket$ must be positive for any $r$. 
Furthermore, as mentioned earlier, the radius $R$ is close to $a$. 
Thus, we have for $r\gg l_p$
\begin{equation}\lb{Tcon1}
-\bra T^t{}_t (r)\ket=\MO(r^{-2})>0.
\end{equation}
On the other hand, from \Ref{Einstein} and \Ref{R}, we obtain $\bra T^\mu{}_\mu\ket=\bra T^t{}_t\ket+\bra T^r{}_r\ket+2\bra T^\th{}_\th\ket=\MO(r^{0})$. 
Therefore, from \Ref{Tcon1}, we have for $r\gg l_p$
\begin{equation}\lb{Tcon2}
\bra T^r{}_r\ket~{\rm or}~\bra T^\th{}_\th\ket=\MO(r^{0}).
\end{equation}

\section{A class of solutions }\lb{s:solve}
\subsection{Derivation}
We solve \Ref{trace} for $l_p\ll r <R$ to find solutions that satisfy \Ref{R}. 
By substituting \Ref{AB}, we can check that 
\Ref{trace} is a non-linear differential equation for $A(r)$ and $B(r)$,
and all the coefficients are only powers of $r$ (see \Ref{eqA1} for the explicit form). 
Therefore, it is natural to consider the power series solution by $\f{1}{r}$.
We set its leading behavior as 
\begin{equation}\lb{AB0}
A(r)=A_0 r^n+\cdots,~~B(r)=B_0 r^m+\cdots,
\end{equation}
where $A_0$ and $B_0$ are constants.
Then, the Einstein tensor is given by 
\begin{align}\lb{G}
-G^t{}_t &=\f{1}{r^2}+\f{m-1}{B_0}r^{-2-m} \nn\\
G^r{}_r &=-\f{1}{r^2}-\f{m-1}{B_0}r^{-2-m}+\f{A_0}{B_0}nr^{-2-m+n} \nn\\
G^\th{}_\th &=\f{m(m-1)}{2B_0}r^{-2-m}-\f{n(3m-2n)A_0}{4B_0}r^{-2-m+n}\nn\\
&~~~+\f{n^2A_0^2}{4B_0}r^{-2-m+2n}.
\end{align}

Using the conditions \Ref{Tcon1} and \Ref{Tcon2} with the Einstein equation \Ref{Einstein}, we can narrow down the possible values of $(n,m)$.  
From \Ref{Tcon1} and $G^t{}_t$ in \Ref{G}, we obtain
\begin{equation}\lb{m_con}
m\geq0.
\end{equation}
We further get the following constraint from \Ref{Tcon2}.
In $G^r{}_r$ and $G^\th{}_\th$ in \Ref{G}, the terms of $r^{-2-m}$ cannot be $\MO(1)$ because of \Ref{m_con}.
Therefore, either $r^{-2-m+n}$ or $r^{-2-m+2n}$ should be $\MO(1)$ to satisfy \Ref{Tcon2}, which holds
only when $n>0$. 
Then, $r^{-2-m+2n}$ is the largest and must be $\MO(1)$. 
This and \Ref{m_con} mean
\begin{equation}\lb{n_con}
m=2n-2~~{\rm and}~~n\geq1.
\end{equation}

So far, we have not used \Ref{trace} yet, 
except for the estimation \Ref{R} and the fact that the power series solution in $\f{1}{r}$ looks natural.
When we calculate both sides of \Ref{trace} for \Ref{AB0} with \Ref{n_con},
the leading terms of both sides are $\MO(r^0)$, and we have (see \Ref{eqA2} for the full form)
\begin{equation}\lb{trace_leading}
\f{n^2A_0^2}{2B_0}+\cdots=8\pi l_p^2 c_W \f{n^4A_0^4}{12B_0^2}+\cdots.
\end{equation}
Thus, for any $n\geq1$, we obtain the relation 
\begin{equation}\lb{AB_sol}
B_0=\f{4\pi l_p^2 c_W n^2 }{3}A_0^2.
\end{equation}

If we assume that the radial pressure $\bra T^r{}_r\ket$ is positive,
we get another constraint on $n$.
In fact, from \Ref{n_con}, \Ref{G} becomes 
\begin{align}\lb{G2}
-G^t{}_t &=\f{1}{r^2}+\f{2n-3}{B_0}r^{-2n}, \nn\\
G^r{}_r &=-\f{1}{r^2}+\f{A_0}{B_0}nr^{-n}+\MO(r^{-2n}),\nn\\
G^\th{}_\th &=\f{n^2A_0^2}{4B_0} + \MO(r^{-n}).
\end{align}
From the second equation,  
for $G^r{}_r=8\pi G\bra T^r{}_r\ket$ to be positive,
we need $n\leq2$.

Thus, we obtain a class of solutions satisfying \Ref{trace} and \Ref{R}, 
whose asymptotic behavior for $r\gg l_p$ is given by
\begin{equation}\lb{sol}
ds^2=-\f{e^{A_0 r^n}}{B_0r^{2n-2}}dt^2+B_0 r^{2n-2}dr^2 +r^2 d\Omega^2.
\end{equation}
Here, $A_0$ and $B_0$ are related by \Ref{AB_sol}, and $n$ satisfies 
\begin{equation}\lb{n_con3}
1\leq n \leq2. 
\end{equation}
Note that the single equation \Ref{trace} together with the asymptotic behavior \Ref{Tcon1} 
has determined the asymptotic functional forms of $A(r)$ and $B(r)$ as \Ref{sol}, 
except for one coefficient $A_0$.

\subsection{Some features of the metrics}
We here point out three salient features of the metric \Ref{sol} 
for any $n$ within the range \Ref{n_con3}.
The first is that 
it has an almost Planckian tangential pressure, 
as can be seen from \Ref{Einstein}, \Ref{AB_sol}, and \Ref{G2}:
\begin{equation}\lb{p}
\bra T^\th{}_\th\ket=\f{1}{8\pi G}\f{3}{16\pi c_W l_p^2}.
\end{equation}
This breaks the dominant energy condition \ci{Poisson} and makes the interior anisotropic $(\bra T^\th{}_\th\ket\gg\bra T^r{}_r\ket)$, 
hence the interior is not fluid. 
Nonetheless, this pressure is consistent with the 4D Weyl anomaly, 
and in fact, it balances the gravity acting on the object. 
In Sec.\ref{s:Einstein}, we will see this large pressure in a field-theoretic calculation.  

The second is that \Ref{sol} is approximately a product of $AdS_2$ and warped $S^2$ for $r\gg l_p$. 
In general, the 2D metric $ds^2=-f(r)dt^2+h(r)dr^2$ 
is $AdS_2$ of radius $L$ 
if the following condition holds \ci{foot:AdS}:
\begin{equation}\lb{AdS}
\sqrt{h(r)}=\f{L}{2}|\p_r \log f(r)|.
\end{equation}
For $f(r)=\f{e^{A_0 r^n}}{B(r)}$ and 
 $h(r)=B(r)=B_0 r^{2n-2}$, \Ref{AdS} is satisfied 
if we neglect the contribution from $\f{1}{B(r)}$ in $f(r)$ for $r\gg l_p$. 
Here, from \Ref{AB_sol} we have
\begin{equation}\lb{AdS_L}
L=\f{2\sqrt{B_0}}{n A_0}=\sqrt{\f{16\pi c_W}{3}}l_p. 
\end{equation}
Thus, to a good approximation for $r\gg l_p$, the $t-r$ part of \Ref{sol} is $AdS_2$ of radius $L=\MO(\sqrt{N}l_p)$, 
and the whole geometry is a warped product of $AdS_2$ and $S^2$.
Indeed, \Ref{sol} has the curvatures  
\begin{equation}\lb{RRR}
R=-\f{2}{L^2},~~R_{\mu\nu}R^{\mu\nu}=\f{2}{L^4},~~R_{\mu\nu\a\b}R^{\mu\nu\a\b}=\f{4}{L^4}
\end{equation}
for $r\gg l_p$ except for $\MO(r^{-n})$ corrections, which is consistent with \Ref{R} \ci{foot:boxR}. 

The third is that the interior metric \Ref{sol} is smoothly connected to the exterior one \Ref{Sch} at $r=R$. 
In fact, we can determine the position of the surface, $r=R$, 
by requiring that each component of \Ref{sol} and \Ref{Sch} agree at $r=R$. 
For $g_{rr}$, we have
\begin{align}\lb{R_der}
  g_{rr}(R)&=B_0 R^{2n-2}=\f{R}{R-a}\nn\\
\Rightarrow R&\approx a\l(1+\f{a^{-2n+2}}{B_0}\r),
\end{align}
where we have used $B_0=\MO(1)$, $a,R\gg l_p$, and \Ref{n_con3}. 
Then, $g_{tt}$ becomes continuous at $r=R$ if we redefine $t$ to $e^{-\f{1}{2}A_0 R^n}t$ in \Ref{sol}
so that 
\begin{equation}\lb{redef_t}
A(r)=A_0r^n\to A_0(r^n-R^n).
\end{equation}
This makes the time coordinate $t$ common inside and outside. 
Finally, $g_{\th\th}$ and $g_{\phi\phi}$ are connected trivially. 

Because of $B_0>0$ from \Ref{AB_sol}, 
\Ref{R_der} shows that $R>a$ and the object has no horizon, as mentioned before. 
Furthermore, \Ref{R_der} tells that 
the object for $n=2$ is the most compact in the range \Ref{n_con3} 
and can be expected to be the black hole. 
In the next section, we will show that this is the case. 
Therefore, the metric \Ref{sol} for $1\leq n <2$ should represent 
the interior of some objects whose size is larger than the radius of the black hole. 
It may be useful for analyzing ultra-dense stars.

\section{Interior metric of the black hole}\lb{s:n=2}
\subsection{Consistency with Hawking radiation}\lb{s:conH}
We show that the choice of $n=2$ is consistent with Hawking radiation. 
We first observe that, when the object is stationary in the heat bath, 
it should continuously emit some radiation to balance the radiation from the bath, as in Fig.\ref{f:BH}.
Now, suppose that we move the object from the bath to the vacuum region. 
Then it will continue to emit the radiation, 
but there will be no radiation received from the bath. 
So, the object will lose the energy and the radius of its surface $R(t)$ will decrease as a
function of time $t$.
This is nothing but the evaporation of a black hole. 

Because most of the total mass $M(t)$ of the system comes from inside the surface, 
the energy flux measured at $r\gg a$ is given from \Ref{mass} by
\begin{equation}\lb{J}
J(t):=-\f{dM(t)}{dt} =- \f{dR(t)}{dt} 4\pi R(t)^2(-\bra T^t{}_t(R(t))\ket).
\end{equation}
Here, we can show that for $1<n\leq2$, the surface shrinks approximately at the speed of light:
\begin{equation}\lb{eomR}
\f{dR(t)}{dt} =-\f{1}{B(R(t))},
\end{equation}
and for $n=1$ a positive coefficient less than or equal to 1 appears in the numerator 
(see Appendix \ref{A:surface} for details).
Then, using \Ref{Tcon1}, \Ref{J}, \Ref{eomR}, $B(r)=B_0r^{2n-2}$, and $R\sim a$, 
we have $J(t)\sim a(t)^{-2n+2}$. 
For $n=2$, this agrees with the standard result of Hawking radiation, $J(t) \sim a(t)^{-2}$ \ci{Hawking,BD}.
Thus, we can conclude that the metric \Ref{sol} with $n=2$ describes the interior of the black hole.
In the following, we assume that $n=2$.

\subsection{Metric with $n=2$}\lb{s:metric2}
Let's discuss the physical meaning of $A_0$ and $B_0$. We parametrize them as
$A_0=\f{1}{2\s \eta},B_0=\f{1}{2\s}$, which means
\begin{equation}\lb{AB_sol2}
A(r)=\f{r^2}{2\s \eta},~~B(r)=\f{r^2}{2\s},~~\s\equiv\f{8\pi l_p^2 c_W}{3\eta^2}.
\end{equation}
Here, $\eta$ is a parameter and the third equation comes from \Ref{AB_sol}.
Then, from \Ref{Einstein} and \Ref{G2}, we have 
\begin{align}\lb{Tn2}
-\bra T^t{}_t\ket&=-\f{1}{8\pi G}G^t{}_t =\f{1}{8\pi Gr^2}, \nn\\
\bra T^r{}_r\ket&=\f{1}{8\pi G}G^r{}_r=\f{2-\eta}{\eta}\f{1}{8\pi Gr^2}.
\end{align}
Furthermore, from \Ref{J}, \Ref{eomR}, \Ref{AB_sol2}, \Ref{Tn2} and $R(t)\approx a(t)$, we have
\begin{equation}\lb{J2}
J(t)=\f{\s}{G a(t)^2}. 
\end{equation}
This is Hawking-like radiation of temperature $\sim \f{\hbar}{a(t)}$, 
where $\s$ is proportional to the Stefan-Boltzmann constant. 
Here, $\eta$ is an analogue of the graybody factor \ci{Landau_SM} 
and must satisfy
\begin{equation}\lb{eta_con}
0<\eta<2
\end{equation}
in order for the condition $\bra T^r{}_r\ket>0$ to hold. 
Note that $\s$ is proportional to $c_W$, 
which is consistent with the fact that $c_W>0$ \ci{Duff,cW}. 
In Sec.\ref{s:Hawking}, we will see 
that these results are consistent with the analysis of the Hawking-like radiation by ordinary field theory.

From \Ref{R_der} and \Ref{AB_sol2}, the surface is located at 
\begin{equation}\lb{Ra}
r=a+\f{2\s}{a}\equiv R(a)
\end{equation}
for $a\gg l_p$ \ci{foot:smooth}. 
Here, the proper distance between $r=a$ and $r=R(a)$ is estimated as 
\begin{equation}\lb{proper_dis}
\sqrt{g_{rr}}|_{r=R(a)}\f{2\s}{a} =\sqrt{\f{R(a)^2}{2\s}}\f{2\s}{a}\approx \sqrt{2\s}\sim\MO(\sqrt{N}l_p). 
\end{equation}
For large $N$, this is considerably larger than $l_p$, and it makes sense to distinguish 
between these two radii in ordinary field theory. 
Therefore, we can conclude again that this object has no horizon \ci{KMY,KY1,KY2,KY3,KY4}.
It is worth noting that \Ref{Ra} coincides with the location of the ``brickwall" \ci{foot:brick}.

After the redefinition of $t$ \Ref{redef_t}, 
the interior metric, \Ref{AB} with \Ref{AB_sol2}, becomes \ci{KMY,KY1,KY2,KY3,KY4}
\begin{equation}\lb{interior}
ds^2=-\f{2\s}{r^2}e^{-\f{R(a)^2-r^2}{2\s\eta}}dt^2+\f{r^2}{2\s}dr^2+r^2d\Omega^2. 
\end{equation}
This indicates that the redshift is very large deep below the surface. 
When we focus on a point $r=R(a)-\D r$ near the surface, 
the proper time at the point for a given time scale $\D t$ becomes 
$\D \tau \equiv \sqrt{-g_{tt}(r)}\D t\approx \f{\sqrt{2\s}}{a}e^{-\f{a\D r}{2\s\eta}}\D t$. 
On the other hand, 
the time scale $\tau_r$ that characterizes the speed of the time evolution at $r$ 
is given by the inverse of the proper acceleration required to stay at $r$: 
\begin{align}\lb{tau_r}
\tau_r&=(g_{\mu\nu}\a^\mu \a^\nu)^{-\f{1}{2}} \nn\\
&=\l(\f{\p_r \log \sqrt{-g_{tt}(r)}}{ \sqrt{g_{rr}(r)}}\r)^{-1}\approx\sqrt{2\s\eta^2},
\end{align}
where $\a^\mu \equiv u^\nu\N_\nu u^\mu$ and $u^\mu\p_\mu=(-g_{tt}(r))^{-\f{1}{2}}\p_t$, 
and \Ref{interior} has been applied.
If $\D \tau < \tau_r$, that is, 
\begin{equation}\lb{freez_t}
\D r > \f{2\s\eta}{a} \log \f{\D t}{\eta a},
\end{equation}
then the time at $r$ flows slowly. 
Therefore, for $\Delta t = \MO(\f{a^{k+1}}{l_p^k}) $ with $k>2$, 
time flows only in the narrow region near the surface $\D r < \f{2k \s\eta}{a} \log \f{a}{l_p}$, 
and the region below is frozen and almost completely unaffected by the outside. 
This justifies our initial assumption that the interior metric $g^{(in)}_{\mu\nu}$ does not
depend on the total mass of the black hole.
Conversely, a signal emanating from a point in the deep region (e.g. $r=\f{a}{2}$) 
will not emerge until a long time $\Delta t \sim \MO(a e^{a^2/l_p^2})$ has elapsed in the heat bath, 
or until the surface has shrunk to that point due to the evaporation in the vacuum.

Note that the interior metric \Ref{interior} is a non-perturbative solution for $\hbar$ 
to the trace part equation \Ref{trace}. 
In fact, we cannot take the limit $\hbar\to0$ in \Ref{interior} and \Ref{RRR} 
because $\s \propto \hbar c_W$ from \Ref{AB_sol2}. 
This result comes from the full 4D dynamics, and the metric contains non-perturbatively the backreaction 
from quantum fluctuation of all modes with arbitrary angular momentum. 
Therefore, the picture here is very different from that of 2D dynamics models \ci{CGHS,RST}. 

\section{Semi-classical Einstein equation}\lb{s:Einstein}
\subsection{Renormalized energy-momentum tensor}\lb{s:EMT}
As we have seen, the metric is determined from the trace part \Ref{trace} 
of the semi-classical Einstein equation, except for the parameter $\eta$. 
It is natural to think that $\eta$ can be  
determined by examining the complete equation \Ref{Einstein}. 
This requires a complicated calculation 
to find the renormalized energy-momentum tensor $\bra \psi| T_{\mu\nu} |\psi \ket$. 
In \ci{KY4}, we evaluated directly $\bra \psi| T_{\mu\nu} |\psi \ket$ for $N$ massless free scalar fields 
and 
determined the self-consistent value of $\eta$. 
In the following, we briefly summarize the results of \ci{KY4}.  

We start with 
the mode expansion of the scalar fields in the background metric \Ref{interior}.
It can be seen that s-waves (with energy $\sim \f{\hbar}{a}$) coming from the outside can enter the black hole of mass $\f{a}{2G}$,
but modes with non-zero angular momentum appear 
as bound modes inside the black hole. 
We then find that the state $|\psi \ket$ suitable for \Ref{Einstein} is the one in which the bound modes are 
in the ground state and the s-waves are 
in a state properly excited to represent the radiation being exchanged with the heat bath.

Therefore, we can write the energy-momentum tensor as 
\begin{equation}\lb{EMT_semi}
\bra\psi| T_{\mu\nu}|\psi\ket=\bra0| T_{\mu\nu}|0\ket+T_{\mu\nu}^{(\psi)},
\end{equation}
where $|0\ket$ is the ground state of all modes in the metric \Ref{interior},  
and $T_{\mu\nu}^{(\psi)}$ is the contribution from the excitation of the s-waves. 
In fact, $\bra0| T_{\mu\nu}|0\ket$ can be evaluated as a well-defined finite quantity
by introducing appropriate counterterms through dimensional regularization.
On the other hand, $T_{\mu\nu}^{(\psi)}$ is determined by the condition that 
the radiation from the bath is balanced by the Hawking-like radiation. 
In this way, under the background metric \Ref{interior}, 
we can calculate $\bra\psi| T_{\mu\nu}|\psi\ket$, where $\eta$ appears in a non-trivial form.
Finally, by comparing the both sides of \Ref{Einstein},
we can determine the self-consistent value of $\eta$ so that \Ref{eta_con} is satisfied.  

In the calculation above, 
by integrating modes over both frequency $\omega$ and angular momentum $l$, 
we obtain $\bra 0| T^\th{}_\th |0\ket=\MO(r^0)$ like \Ref{p}, 
which means that the large tangential pressure originates from the vacuum fluctuations of the bound modes with various $l$. 
Furthermore, this pressure induces the Hawking-like radiation with arbitrary angular momentum, 
along with negative energy flow, to each point inside the object \ci{KY1,KY2,KY4}.  

In this way, we can check that 
as a result of 4D dynamics of the fields, the metric \Ref{interior} satisfies the semi-classical Einstein equation \Ref{Einstein} 
self-consistently. 

Note here that 
the area law of entropy can be explained by states $\{|\psi\ket\}$ satisfying \Ref{Einstein} and \Ref{EMT_semi} \ci{KY4}. 
In fact, by counting their number, we can evaluate 
the entropy per unit proper radial distance as 
\begin{equation}\lb{entropy}
s=\f{2\pi\sqrt{2\s}}{l_p^2}.
\end{equation} 
Integrating it inside the object, we have $S=\int^{R(a)}_0 dr \sqrt{g_{rr}(r)}s\approx \f{A}{4l_p^2}$. 

\Ref{entropy} through $\s \sim N l_p^2$ means that 
$\MO(\sqrt{N})$ bits of information are packed per the proper radial length of one Planck length. 
To understand this intuitively, let's consider Bekenstein's thought experiment \ci{Bekenstein} 
based on the picture of the black hole consisting of layers like a brick wall, as we have discussed. 
At each stage of slow formation where the size of the object is $\sim r$ and the temperature is $T\sim \f{\hbar}{r}$, 
each wave with energy $\e \sim T\sim \f{\hbar}{r}$ forms a layer 
with one bit of information that the wave may or may not enter the black hole, 
since the wavelength $\lambda \sim \f{\hbar}{\e}\sim r$ 
is almost the same as the size of the black hole \ci{Bekenstein}. 
The layer at $r$ has the width $\D r=2G\e \sim \f{l_p^2}{r}$, 
whose proper length is $\D l=\sqrt{g_{rr}}\D r \sim \f{l_p}{\sqrt{N}}$ 
from $g_{rr}=\f{r^2}{2\s}$. 
Then, we can estimate the entropy density $s$ as $s\sim \f{1~{\rm bit}}{\D l}\sim\f{\sqrt{N}}{l_p}$, 
which agrees with \Ref{entropy}.

\subsection{Hawking-like radiation}\lb{s:Hawking}
In this subsection, 
we consider the metric \Ref{interior} as a background 
and analyze the time evolution of quantum matter fields   
to demonstrate that the Hawking-like radiation occurs self-consistently to \Ref{J2}. 
In particular, we see that it is generated in a region of width $\D r \sim \f{\s}{a}$ below the surface.

We first set up the background geometry. 
As seen in Sec.\ref{s:conH}, when taken out of the heat bath, 
the object starts to evaporate. 
Then, because the interior region is frozen due to the large redshift, 
it is still described by the stationary metric \Ref{interior}, 
except that $R(a)= a+\f{2\s}{a}$ is replaced by $R(a(t))= a(t)+\f{2\s}{a(t)}$: 
\begin{equation}\lb{interiort}
ds^2=-\f{2\s}{r^2}e^{-\f{R(a(t))^2-r^2}{2\s\eta}}dt^2+\f{r^2}{2\s}dr^2+r^2d\Omega^2.
\end{equation}
Here, $a(t)$ decreases (from \Ref{eomR} and $R(a(t))\approx a(t)$) as 
\begin{equation}\lb{dat}
\f{da(t)}{dt}=-\f{2\s}{a(t)^2}. 
\end{equation}

For the sake of argument, we express this metric in a $(u,r)$ coordinate \ci{KMY,KY2,KY3,KY4}: 
\begin{equation}\lb{interior2}
ds^2=-\f{2\s}{r^2}e^{-\f{R(a(u))^2-r^2}{2\s\eta}}du^2-2e^{-\f{R(a(u))^2-r^2}{4\s\eta}}dudr+r^2d\Omega^2,
\end{equation}
with $R(a(u))=a(u)+\f{2\s}{a(u)}$ and 
\begin{equation}\lb{da}
\f{da(u)}{du}=-\f{\s}{a(u)^2},
\end{equation}
where we have defined $u$ by 
$e^{-\f{R(a(u))^2}{4\s\eta}}du\equiv e^{-\f{R(a(t))^2}{4\s\eta}}dt - \f{r^2}{2\s}e^{-\f{r^2}{4\s\eta}}dr$ \ci{foot:u_derive}. 
Note that for $a(u)=$const., the $u$ coordinate becomes the outgoing Eddington-Finkelstein coordinate, 
since \Ref{interior2} connects to the Schwarzschild metric \Ref{Sch} in that coordinate at $r=R(a)$.

\begin{figure}[h]
\begin{center}
\includegraphics*[scale=0.1]{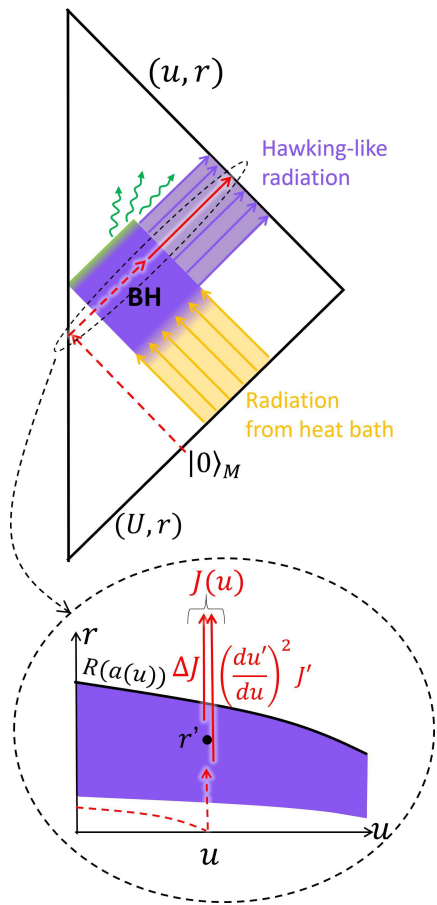}
\caption{Above: the Penrose diagram for the formation and evaporation of the slowly-formed black hole. 
The center violet part is the semi-classical region described by \Ref{interior2}, 
while the small green one is assumed as a quantum-gravity region with mass $\sim m_p$ and lifetime $\sim l_p$. 
$U$ and $u$ are the outgoing null time coordinates in the past and future null infinity, respectively. 
Below: the conservation law of the energy flux of the particles created by the time evolution of quantum fields.}
\label{f:Pen}
\end{center}
\end{figure}

Now, we study the propagation of the matter fields in the background geometry \Ref{interior2}. 
To do that, we goes back to the formation process of the object 
and consider the time evolution of the fields during the formation \ci{Hawking}. 
Suppose that they start from the initial flat space in the Minkowski vacuum state $|0\ket_M$, 
pass the center at $r=0$, propagate through the interior metric \Ref{interior2}, 
come out of the surface, and go to infinity (see the above part of Fig.\ref{f:Pen}). 
During this process, the redshift the fields feel changes in time, 
leading to particle creation. 

Then, using the energy conservation, 
the energy flux observed at $r\gg a$ and $u=$const. can be expressed as 
\begin{align}\lb{J_gene}
J(u)&\equiv 4\pi r^2 {}_M\bra0| T_{uu}|0\ket_M|_{r\gg a}\nn\\
 &=\D J + \l(\f{du'}{du}\r)^2 J'.
\end{align}
See the below part of Fig.\ref{f:Pen}. 
Here, $u'$ is an arbitrary local-time coordinate at a point $r=r'$ on a $u$-constant line below the surface, 
and $\D J$ and $J'$ express the energy fluxes generated in the region $r'\leq r \leq R(a(u))$ 
and $r\leq r'$, respectively. 
Therefore, \Ref{J_gene} means that 
$J$ is the sum of $\D J$, which comes directly, 
and $J'$, which is weakened by the redshift $\l(\f{du'}{du}\r)^2$. 
If the redshift is sufficiently strong, only $\D J$ will contribute to $J$. 
For example, when we consider a timelike hypersurface with $r={\rm const.}=r'$, 
$u'$ is defined as the proper time coordinate at $r'$ in \Ref{interior2}: 
\begin{equation}\lb{u'}
du'\equiv \f{\sqrt{2\s}}{r'} e^{-\f{R(a(u))^2-r'^2}{4\s\eta}} du,
\end{equation}
which means that $\f{du'}{du}$ is exponentially small for $R(a(u))-r'\gg \f{\s}{a}$. 
Thus, the energy flux reaching $r\gg a$ comes from the region with width $\D r \sim \f{\s}{a}$ below the surface. 

For an explicit demonstration, 
we consider s-waves of $N$ massless free scalar fields 
in the eikonal approximation \ci{KMY,KY2} (or 2D approximation \ci{KY4,Strom}). 
Then, we can evaluate the energy flux as 
\begin{align}\lb{J_form}
J(u)&=\f{\hbar N}{16\pi}\{u,U\}\nn\\
&=\f{\hbar N}{16\pi}\l(\{u,u'\}+ \l(\f{du'}{du}\r)^2\{u',U\} \r).
\end{align}
Here, $\{x,y\}\equiv \f{\ddot y^2}{\dot y^2}-\f{2}{3}\f{\dddot y}{\dot y}$ is the Schwarzian derivative 
for $y=y(x)$, $U$ is the outgoing null time in the flat space before the formation, 
and we have used a formula $\{z,x\}=\{z,y\}+\l(\f{dy}{dz}\r)^2\{y,x\}$ \ci{CFTbook}. 
This takes the same form as the general relation of the energy conservation \Ref{J_gene}.
To evaluate \Ref{J_form}, we first note that 
\begin{equation}\lb{J_eva1}
\{u,u'\} = \f{1}{3}\l(\f{d\xi}{du}\r)^2-\f{2}{3}\f{d^2\xi}{du^2},~\xi\equiv \log \f{du'}{du}.
\end{equation}
From \Ref{u'}, we have
\begin{align}\lb{J_eva2}
\xi &\approx \f{-R(a(u))^2+r'^2}{4\s\eta} \nn \\
\Rightarrow \f{d\xi}{du} &\approx -\f{a(u)}{2\s\eta}\f{da(u)}{du} \nonumber \\
 &= \f{1}{2\eta a(u)},
\end{align}
where we used $R(a(u))\approx a(u)$ and \Ref{da}.
For a deep point $r'$ satisfying  $R(a(u))-r'\gg \f{\s}{a}$, 
we can drop the second term in \Ref{J_form} unless $\{u',U\}$ is exponentially large. 
Thus, combining \Ref{J_eva1} and \Ref{J_eva2}, 
\Ref{J_form} becomes 
\begin{equation}\lb{J_form2}
J(u)=\f{\hbar N}{192\pi \eta^2 a(u)^2} + \MO(a^{-4}).
\end{equation}
This reproduces self-consistently the Hawking-like radiation with the intensity 
\begin{equation}\lb{sigma_s}
\s_{s}=\f{N l_p^2}{96\pi \eta^2},
\end{equation}
which is obtained by comparing \Ref{J_form2} and $J(u)=\f{\s}{2G a^2}$ from \Ref{da} \ci{foot:sigma1,foot:sigma2}.

The point is that, 
because of the exponentially large redshift, 
the Hawking-like radiation \Ref{J_form2} is determined by the geometry \Ref{interior2} of 
the region with width $\D r\sim\f{\s}{a}$ near the surface,
and the result is independent of the detail of the deeper region \ci{foot:differ}.
In this sense, the Hawking-like radiation will occur universally for a general black hole 
as long as its near-surface region has approximately the same metric as \Ref{interior2}. 

We conclude that, although the object has no horizon,
the Hawking-like radiation occurs self-consistently. 
Note that, using \Ref{u'} and \Ref{da}, we can also obtain a Planck-like distribution of 
the time-dependent temperature $T=\f{\hbar}{4\pi a(u)}$ \ci{KMY,KY2}.

\section{Conclusion and Discussion}\lb{s:Con}
We have considered the interior metric $g_{\mu\nu}^{(in)}$ of 
a black hole that has grown slowly in a heat bath. 
By assuming the $a$-independence of $g_{\mu\nu}^{(in)}$, 
the positivity of $\bra T^r{}_r\ket$, and the evaporation of black holes, 
the trace part \Ref{trace} of the semi-classical Einstein equation 
determines $g_{\mu\nu}^{(in)}$ \Ref{interior}, which is a non-perturbative solution for $\hbar$. 
This is smoothly connected to the Schwarzschild metric \Ref{Sch} at $r=R(a)$, \Ref{Ra}, 
implying that the black hole is a dense object with a surface located slightly 
outside the Schwarzschild radius $a$, as in Fig.\ref{f:BH}. 
In summary, the metric for the entire region is given by 
\begin{equation}\lb{fullmetric}
ds^2=\begin{cases}
 -\f{2\s}{r^2}e^{-\f{R(a)^2-r^2}{2\s\eta}}dt^2+\f{r^2}{2\s}dr^2+r^2d\Omega^2,~{\rm for}~r\leq R(a),\\
 - \f{r-a}{r}d t^2 + \f{r}{r-a}d r^2 + r^2 d \Omega^2,~~{\rm for}~~R(a)\leq r.
\end{cases}
\end{equation}

When taken out of the heat bath, 
the object evaporates from the outside, like peeling an onion. 
The interior metric is given by \Ref{interior2} with \Ref{da} 
while the exterior one may be different from \Ref{Sch}. 
For example, we can use the Vaidya metric \ci{Poisson} as a first approximation. 
Then, we have
\begin{equation}\lb{fullmetric2}
ds^2=\begin{cases}
 -\f{2\s}{r^2}e^{-\f{R(a(u))^2-r^2}{2\s\eta}}du^2-2e^{-\f{R(a(u))^2-r^2}{4\s\eta}}dudr+r^2d\Omega^2\\
 ~~~~~~~~~~~~~~~~~~~~~~~~~~~~~~~~~~~~~~~~~~~~~~{\rm for}~r\leq R(a(u)),\\
 -\l(1-\f{a(u)}{r}\r)du^2-2dudr + r^2 d \Omega^2~{\rm for}~R(a(u))\leq r.
\end{cases}
\end{equation}
This describes the evaporating black hole in the vacuum region, 
except for the final stage of the evaporation. 
It is natural to think of a small black hole with $a\sim l_p$ 
as a state of string theory that decays in time $\D u \sim l_p$.
Then, the Penrose diagram for the whole process of the black-hole formation and evaporation 
is identical to the flat spacetime (see the above part of Fig.\ref{f:Pen}) \ci{KMY,KY2,KY4}.

Also, we have shown that the time evolution of quantum matter fields in the background metric 
reproduces the Hawking-like radiation in a self-consistent manner. 

This picture unifies two controversial views on the quantum nature of black holes: 
one is that the energy density near $r=a$ is $\MO(a^{-4})$, which is small (i.e., ``uneventful") \ci{DFU,Fulling,Howard}, 
and the other is that many excited states near $r=a$ form a structure like ``brickwall" \ci{brick}, ``fuzzball" \ci{Mathur}, or ``firewall" \ci{AMPS}, 
which have a large backreaction to the geometry (i.e., ``eventful"). 
The solution obtained above shows that the inside of the surface is eventful while the outside is uneventful.

The new point compared to the previous works \ci{KMY,KY1,KY2,KY3,KY4} is 
the direct analysis of the 4D Weyl anomaly. 
That is, we have just solved \Ref{trace} to find the class of solutions \Ref{sol}, 
and used the consistency with Hawking radiation to find \Ref{AB_sol2} as the interior metric of the black hole. 
In fact, the metric is uniquely obtained, 
independently of the state $|\psi \ket$ and the details of the exterior metric \ci{foot:universal}. 
This should provide the robustness of our picture of the black hole. 

Furthermore, from a technical standpoint, this method will be very useful in the future. 
Since the trace part equation \Ref{trace} is valid for any spacetime and any state, 
it will be a key equation for investigating time-dependent or non-spherically symmetric geometries. 

For example, considering a more general (not slow but fast) collapse of matter, 
the outermost regions are expected to approach the universal metric \Ref{AB_sol2} in a short time, 
about the ``scrambling time" \ci{Sekino}, 
while the deep regions are expected to retain the details of the initial distribution of the matter 
for a long time due to large redshifts \ci{KY2,Ho3}. 
We plan to study this conjecture in the near future 
by using \Ref{trace} to analyze the time evolution of the fast formation 
and the stability of the metric \Ref{interior} for the time scale $\Delta t >\MO(ae^{a^2/l_p^2})$ \ci{KYnew}.
If it is true, a general black hole should be considered 
as a perturbation from the slowly-formed one. 

Another application of \Ref{trace} is 
to study the interior of an electrically-charged black hole and a rotating one. 
In fact, the former one can be obtained under the assumption 
that electric charges move quickly outward 
due to the repulsive force in the slow formation process inside the heat bath \ci{KY2}. 
As the result, the charged black hole is like a spherical condenser: 
the interior metric is the same as \Ref{AB} with \Ref{AB_sol2}, 
the charges are distributed only on the surface whose location is slightly different from \Ref{Ra}, 
and the exterior metric is given by the Reissner-Nordstr\"{o}m metric. 
On the other hand, it is less trivial to construct the interior metric of the rotating one 
because of the non-spherical symmetry, 
although its surface can be identified \ci{KY2}. 
In future, we want to address this problem by employing \Ref{trace}. 

Finally, an important aspect of this picture is that matter is distributed in the interior 
and the Hawking-like radiation occurs in the same place, as seen in Sec.\ref{s:Einstein}. 
They can then interact with each other at the almost Planckian energy scale 
and exchange information directly \ci{KY2}. 
Thus, the information of the initial collapsing matter should be  
reflected in the Hawking-like radiation, 
and the whole black hole can be thought of as just one long-lived resonance.
It is an interesting problem to figure out how all the information that the matter had 
comes back as the radiation.

\section*{Acknowledgements}
We thank P.M. Ho, H. Liao and E. R. Livine for valuable comments. 
H.K. thanks Professor Shin-Nan Yang and his family for their kind support through the Chin-Yu chair professorship. 
H.K. is also partially supported by Japan Society of Promotion of Science (Grants No.20K03970 and No.18H03708), 
by the Ministry of Science and Technology, R.O.C. (MOST 110-2811-M-002-500), and by National Taiwan University.
Y.Y. is partially supported by RIKEN iTHEMS Program and by Japan Society of Promotion of Science (Grants No.21K13929, No.18K13550, and No.17H01148). 
\appendix 
\section{Multi-shell model}\lb{A:model}
For a self-contained discussion, we give a short review of a multi-shell model \ci{KMY,KY3,KY4}, 
which is introduced in Sec.\ref{s:metric}, 
to explain how the $a$-independent interior metric $g_{\mu\nu}^{(in)}$ is obtained.
Note that the main discussion in this paper provides a robustness of the picture of Fig.\ref{f:BH}, 
independently of this model. 

\begin{figure}[h]
\begin{center}
\includegraphics*[scale=0.1]{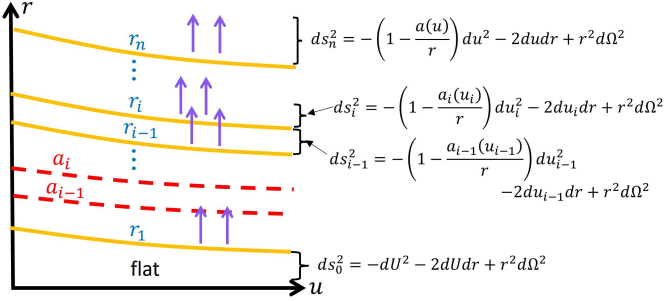}
\caption{A multi-shell model.}
\label{f:model}
\end{center}
\end{figure}

As in Fig.\ref{f:shells}, 
we model radiations coming from the bath as $n(\gg1)$ concentric null thin shells. 
To incorporate the backreaction from the particle creation during the collapse, 
we assume that each shell evaporates as the usual black hole, 
and that the metric just outside each shell is given by the Vaidya metric \ci{Poisson}:
\begin{equation}\lb{model1}
ds^2_i=-\l(1-\f{a_i(u_i)}{r}\r)du_i^2-2du_idr+r^2 d\Omega^2,
\end{equation}
with
\begin{equation}\lb{model2}
\f{da_i(u_i)}{du_i}=-\f{\s}{a_i^2}
\end{equation}
for $i=0,1,2\cdots n$. 
See Fig.\ref{f:model}. 
Here, $\f{\D a_i}{2G}=\f{a_i-a_{i-1}}{2G}$ is the energy of the $i$-th shell, $u_i$ is the local time just above it, 
and $\s$ is the intensity of Hawking-like radiation. 
The total size $a_n$ is $a$, the outer most time $u_n$ is $u$, and the center is flat: 
$u_n=u,~a_n=a;~u_0=U,~a_0=0.$
To connect these coordinates, we use the fact that 
each shell, whose locus is denoted by $r=r_i(u_i)$, moves at the speed of light in both its exterior and interior metrics:
\begin{equation}\lb{model3}
\f{r_i-a_i}{r_i}du_i=-2dr_i=\f{r_i-a_{i-1}}{r_i}du_{i-1}.
\end{equation}
This is equivalent to 
\begin{align}\lb{model4}
\f{dr_i(u_i)}{du_i}&=-\f{r_i(u_i)-a_i(u_i)}{2r_i(u_i)},\\
\lb{model5}
\f{du_i}{du_{i-1}}&=\f{r_i-a_{i-1}}{r_i-a_i}=1+\f{a_i-a_{i-1}}{r_i-a_i}.
\end{align}
Note that for each $i$ this equation system is closed up to the $i$-th one. 
This means that due to the spherical symmetry and the shells being null, 
the $i$-th shell and metric only depend on the region inside it.

From now, we consider the asymptotic behavior of the shells distributed in a continuum way ($\D a_i\to 0$) 
and construct the continuum metric. 

First, we study where $r_i(u_i)$ will approach in the metric \Ref{model1} with \Ref{model2}. 
When $r_i\sim a_i$, we can replace $r_i$ in the denominator of \Ref{model4} by $a_i$, 
set $\D r_i(u_i)\equiv r_i(u_i)-a_i(u_i)$ and obtain 
\begin{equation}\lb{model6}
\f{d \D r_i(u_i)}{d u_i}\approx -\f{\D r_i(u_i)}{2a_i(u_i)}-\f{da_i(u_i)}{du_i}. 
\end{equation}
The first term is negative, which is the effect of the collapse, 
and the second one is positive, which is the effect of the evaporation \Ref{model2}. 
When $\D r_i(u_i)\sim \f{l_p^2}{a_i(u_i)}$, 
the both terms are balanced so that the right-hand side vanishes. 
This means that each shell behaves asymptotically as 
\begin{equation}\lb{model7}
r_i(u_i)\to a_i(u_i)-2a_i(u_i) \f{da_i(u_i)}{da_i}= a_i(u_i)+\f{2\s}{a_i(u_i)}.
\end{equation}
(See \ci{KY2} for the details.) 
It is worth noting that no matter where the shell starts from, 
as long as $r_i$ comes close to $a_i$, 
the shell will behave like \Ref{model7} universally. 

Next, we take a continuum limit $\D a_i\to 0$ and assume that each shell reaches the asymptotic position, $r_i=a_i+\f{2\s}{a_i}$. 
Then, the shells pile up continuously and become a spherical dense object with mass $\f{a}{2G}$, as in Fig.\ref{f:BH}. 
We evaluate the redshift factor $\xi_i\equiv\log \f{dU}{du_i}$, as follows. 
\begin{align}
\xi_i-\xi_{i-1} &=\log \f{\f{dU}{du_i}}{\f{dU}{du_{i-1}}}=-\log\f{du_i}{du_{i-1}} \nn\\
 &=-\log\l(1+\f{a_i-a_{i-1}}{r_i-a_i}\r) \nonumber \\
 &\approx- \f{a_i-a_{i-1}}{r_i-a_i} = -\f{a_i-a_{i-1}}{\f{2\s}{a_i}} \nonumber \\
 &\approx -\f{1}{4\s} (a_i^2-a_{i-1}^2).
\end{align}
Here, at the second line we have used \Ref{model5}; 
at the third line we have considered $\f{a_i-a_{i-1}}{\f{2\s}{a_i}}\ll1$ in the limit $\D a_i\to0$;
at the final line we have approximated $2a_i \approx a_i+a_{i-1}$. 
Using $u_0=U$ and $a_0=0$, 
we obtain
\begin{equation}\lb{model8}
\xi_i =-\f{1}{4\s}a_i^2.
\end{equation}
Note here that, 
thanks to the fact that \Ref{model4} and \Ref{model5} are closed up to the $i$-th, 
\Ref{model8} can be determined only from the information below the $i$-th.

Now, let's construct the continuum metric in the $(U,r)$ coordinate. 
By considering the shell that passes a spacetime point $(U,r)$ inside the object, 
we have at $r=r_i$
\begin{equation}
\f{r_i-a_i}{r_i}=\f{\f{2\s}{a_i}}{r_i}\approx \f{2\s}{r^2},~~
\f{du_i}{dU} =e^{-\xi_i}=e^{\f{a_i^2}{4\s}}\approx e^{\f{r^2}{4\s}}, \nn
\end{equation}
where we have used \Ref{model7} and \Ref{model8}. 
Then, the metric at that point is given by  
\begin{align}\lb{model9}
ds^2&=-\l(1-\f{a_i}{r_i}\r)du_i^2-2du_idr+r^2_i d\Omega^2,\nn\\
 &=-\l(1-\f{a_i}{r_i}\r)\l(\f{du_i}{dU}\r)^2dU^2-2\l(\f{du_i}{dU}\r)dUdr+r^2_i d\Omega^2 \nonumber \\
 &\approx -\f{2\s}{r^2} e^{\f{r^2}{2\s}}dU^2-2 e^{\f{r^2}{4\s}}dUdr+r^2d\Omega^2,
\end{align}
which is independent of the total size $a$. 
This comes essentially from the universal behavior \Ref{model7} 
and the spherical symmetry of the continuously-distributed null shells.

Finally, using $\f{dU}{du}=e^{-\f{a(u)^2}{4\s}}\approx e^{-\f{R(a(u))^2}{4\s}}$ from \Ref{model8} and \Ref{Ra}, 
the metric \Ref{model9} expressed in the $(u,r)$ coordinate agrees with \Ref{interior2} for $\eta=1$. 
By various methods \ci{KMY,KY3,KY4}, we can show that the Hawking-like radiation occurs self-consistently to \Ref{model2}
and determine the self-consistent value of $\s$ (see also Sec.\ref{s:n=2} and Sec.\ref{s:Einstein}). 

\begin{widetext}
\section{The trace part of the semi-classical Einstein equation}\lb{A:trace}
Using the ansatz \Ref{AB}, the trace part \Ref{trace} of the Einstein equation becomes 
\begin{align}\lb{eqA1}
&\frac{A''(r)}{B(r)}-\frac{3 A'(r) B'(r)}{2
   B(r)^2}+\frac{A'(r)^2}{2 B(r)}+\frac{2 A'(r)}{r
   B(r)}-\frac{B''(r)}{B(r)^2}+\frac{2
   B'(r)^2}{B(r)^3}-\frac{4 B'(r)}{r B(r)^2}+\frac{2}{r^2
   B(r)}-\frac{2}{r^2}  \nn\\
 &=\widetilde c_W\l[-\frac{2 A''(r) B''(r)}{3 B(r)^3}+\frac{4 A''(r) B'(r)^2}{3
   B(r)^4}+\frac{4 A''(r) B'(r)}{3 r B(r)^3}-\frac{4
   A''(r)}{3 r^2 B(r)}+\frac{4 A''(r)}{3 r^2
   B(r)^2}+\frac{A''(r)^2}{3 B(r)^2}-\frac{A'(r)^2 B''(r)}{3
   B(r)^3}+\frac{2 A'(r) B''(r)}{3 r B(r)^3}\r.\nn\\
 &  +\frac{2 A'(r)
   B'(r)}{r^2 B(r)^2}-\frac{10 A'(r) B'(r)}{3 r^2
   B(r)^3}-\frac{A'(r)^3 B'(r)}{2 B(r)^3}+\frac{17 A'(r)^2
   B'(r)^2}{12 B(r)^4}+\frac{5 A'(r)^2 B'(r)}{3 r
   B(r)^3}-\frac{2 A'(r) B'(r)^3}{B(r)^5}
   -\frac{10 A'(r) B'(r)^2}{3 r B(r)^4}\nn\\
 &+\frac{A'(r) B'(r)B''(r)}{B(r)^4}
   +\frac{4 A'(r)}{3 r^3 B(r)}-\frac{4
   A'(r)}{3 r^3 B(r)^2}-\frac{2 A'(r)^2}{3 r^2
   B(r)}+\frac{A'(r)^2}{r^2 B(r)^2}+\frac{A'(r)^4}{12
   B(r)^2}-\frac{A'(r)^3}{3 r B(r)^2}-\frac{A'(r) A''(r)
   B'(r)}{B(r)^3}\nn\\
 &+\frac{A'(r)^2 A''(r)}{3 B(r)^2}-\frac{2
   A'(r) A''(r)}{3 r B(r)^2}+\frac{4 B''(r)}{3 r^2
   B(r)^2}-\frac{4 B''(r)}{3 r^2 B(r)^3}+\frac{B''(r)^2}{3
   B(r)^4}-\frac{8 B'(r)}{3 r^3 B(r)^2}+\frac{8 B'(r)}{3 r^3
   B(r)^3}-\frac{8 B'(r)^2}{3 r^2 B(r)^3}+\frac{4
   B'(r)^2}{r^2 B(r)^4}\nn\\
 &\l.+\frac{4 B'(r)^4}{3 B(r)^6}+\frac{8
   B'(r)^3}{3 r B(r)^5}-\frac{4 B'(r)^2 B''(r)}{3
   B(r)^5}-\frac{4 B'(r) B''(r)}{3 r B(r)^4}-\frac{8}{3 r^4
   B(r)}+\frac{4}{3 r^4 B(r)^2}+\frac{4}{3 r^4}\r] 
 -\widetilde a_W \l[-\frac{4 A''(r)}{r^2 B(r)}\r. \nn\\
 &\l.+\frac{4 A''(r)}{r^2
   B(r)^2}+\frac{6 A'(r) B'(r)}{r^2 B(r)^2}-\frac{10 A'(r)
   B'(r)}{r^2 B(r)^3}-\frac{2 A'(r)^2}{r^2 B(r)}
   +\frac{2 A'(r)^2}{r^2 B(r)^2}+\frac{4 B''(r)}{r^2 B(r)^2}-\frac{4
   B''(r)}{r^2 B(r)^3}-\frac{8 B'(r)^2}{r^2 B(r)^3}+\frac{12
   B'(r)^2}{r^2 B(r)^4}\r]\nonumber \\
 &+\widetilde b_W \l[-\frac{A^{(4)}(r)}{B(r)^2}+\frac{9 A^{(3)}(r) B'(r)}{2
   B(r)^3}-\frac{4 A^{(3)}(r)}{r B(r)^2}+\frac{4 A''(r)
   B''(r)}{B(r)^3}-\frac{21 A''(r) B'(r)^2}{2
   B(r)^4}+\frac{11 A''(r) B'(r)}{r
   B(r)^3}-\frac{A''(r)^2}{B(r)^2}\r.\nn\\
 &+\frac{2 B^{(3)}(r)
   A'(r)}{B(r)^3}+\frac{5 A'(r)^2 B''(r)}{4 B(r)^3}+\frac{7
   A'(r) B''(r)}{r B(r)^3}-\frac{3 A'(r) B'(r)}{r^2
   B(r)^3}+\frac{15 A'(r) B'(r)^3}{B(r)^5}-\frac{3 A'(r)^2
   B'(r)^2}{B(r)^4}-\frac{16 A'(r) B'(r)^2}{r
   B(r)^4}\nn\\
 &+\frac{A'(r)^3 B'(r)}{4 B(r)^3}+\frac{2 A'(r)^2
   B'(r)}{r B(r)^3}-\frac{27 A'(r) B'(r) B''(r)}{2
   B(r)^4}-\frac{2 A'(r)}{r^3 B(r)}+\frac{2 A'(r)}{r^3
   B(r)^2}+\frac{A'(r)^2}{r^2 B(r)^2}-\frac{3 A^{(3)}(r)
   A'(r)}{2 B(r)^2}\nn\\
 &+\frac{17 A'(r) A''(r) B'(r)}{4
   B(r)^3}-\frac{A'(r)^2 A''(r)}{2 B(r)^2}-\frac{3 A'(r)
   A''(r)}{r B(r)^2}+\frac{B^{(4)}(r)}{B(r)^3}+\frac{6
   B^{(3)}(r)}{r B(r)^3}+\frac{2 B''(r)}{r^2 B(r)^3}-\frac{6
   B''(r)^2}{B(r)^4}+\frac{4 B'(r)}{r^3 B(r)^2}-\frac{8
   B'(r)}{r^3 B(r)^3}\nn\\
 &\l.-\frac{2 B'(r)^2}{r^2 B(r)^4}-\frac{30
   B'(r)^4}{B(r)^6}+\frac{44 B'(r)^3}{r B(r)^5}-\frac{9
   B^{(3)}(r) B'(r)}{B(r)^4}+\frac{42 B'(r)^2
   B''(r)}{B(r)^5}-\frac{40 B'(r) B''(r)}{r
   B(r)^4}+\frac{4}{r^4 B(r)}-\frac{4}{r^4 B(r)^2}\r]. 
 \end{align}
Here, $A'(r)$ stands for $\f{dA(r)}{dr}$ and so on, 
and $\widetilde c_W\equiv 8\pi l_p^2 c_W$ and so on.
This is a non-linear differential equation for $A(r)$ and $B(r)$ 
where the coefficients of all the terms are only powers of $r$. 

Next, setting $A(r)=A_0r^n$ and $B(r)=B_0 r^{2n-2}$, \Ref{eqA1} becomes
\begin{align}\lb{eqA2}
0=&\frac{A_0^2 n^2}{2 B_0}-\frac{A_0^4 \cw n^4}{12 B_0^2}
+\frac{2 A_0^2 n^2 (\cw-3 \aw)-6 B_0}{3 B_0 r^2}-\frac{4 \cw}{3 r^4}\nn\\
 &+r^{-n} \left(\frac{A_0^3 \cw n^3 (2 n-1)-6 A_0 B_0(n-2) n}{3 B_0^2}+\frac{2 A_0 n (12 \aw (n-1)+3 \bw-4 \cw n+2 \cw)}{3 B_0r^2}\right)\nn\\
 &+r^{-2 n} \l(\frac{6 B_0 \left(2 n^2-7 n+6\right)-A_0^2 n^2
   \left(-6 \aw-3 \bw (n-2) n+\cw \left(8 n^2-6
   n+1\right)\right)}{3 B_0^2}\r.\nn\\
 &\l.-\frac{4 (2 n-1) (6 \aw(n-1)+3 \bw-2 \cw n)}{3 B_0 r^2}\r)\nn\\
  &+r^{-3n}\frac{2 A_0 n \left(-24 \aw (n-1)-3 \bw n \left(5 n^2-16
   n+12\right)+2 \cw n (1-2 n)^2\right)}{3 B_0^2} \nonumber \\
 &+r^{-4n}\frac{4 \left(6 \aw \left(4 n^2-7 n+3\right)+n \left(3 \bw \left(8 n^3-34 n^2+45
   n-18\right)-\cw (1-2 n)^2 n\right)\right)}{3 B_0^2}.
\end{align}
The first two terms are the largest for $r\gg l_p$ and $n\geq1$, giving \Ref{trace_leading}.
\end{widetext}

\section{Motion of the surface}\lb{A:surface}
We derive the equation of motion of the surface \Ref{eomR} for $1\leq n\leq2$. 
We start with the Hamilton-Jacobi equation of a massless particle in the metric \Ref{AB}, 
\begin{align}\lb{HJ1}
0 &=g^{\mu\nu}\p_\mu S \p_\nu S \nn\\
 &=-Be^{-A} (\p_tS)^2 +B^{-1} (\p_rS)^2 +\f{1}{r^2}(\p_\phi S)^2,
\end{align}
where we have chosen $\theta=\f{\pi}{2}$ because of the spherical symmetry. 
Substituting 
\begin{equation}\lb{HJ2}
S=-Et+L\phi+W(r)
\end{equation}
into \Ref{HJ1} and integrating $\f{dW}{dr}$, we obtain 
\begin{equation}\lb{HJ3}
S=-Et+L\phi\pm\int^r dr'\sqrt{B(Be^{-A}E^2-\f{L^2}{r^2})}.
\end{equation}
Then, we take the two derivatives of \Ref{HJ3} with respect to $E$ and to $t$ 
and get the equation of motion in the radial direction: 
\begin{equation}\lb{HJ4}
\f{dr(t)}{dt}=\pm \l.\f{1}{E B^2 e^{-A}}\sqrt{B(Be^{-A}E^2-\f{L^2}{r^2})}\r|_{r=r(t)}.
\end{equation}

Now, we consider the evaporating object in the vacuum region 
and suppose that the shrinking surface consists of massless particles whose radial positions are $r=R(t)$. 
Applying \Ref{sol} to \Ref{HJ4} and using \Ref{redef_t} where $R$ is replaced by $R(t)$, 
the equation of motion of such a particle with $(E,L)$ is given by 
\begin{equation}\lb{HJ5}
\f{dR(t)}{dt}=-\f{C_n}{B(R(t))}
\end{equation} 
with 
\begin{equation}
C_n\equiv \sqrt{1-\f{L^2}{E^2 R(t)^2 B(R(t))}}.
\end{equation}
Because the particles have come from the heat bath to the object with size $\sim R$,  
their angular momentum can be estimated as 
$L\sim({\rm impact~parameter})\times ({\rm energy})\lesssim R E$. 
Therefore, using $B(r)=B_0r^{2n-2}$, $R\sim a$, and the assumption that 
the length scale $B_0$ can have is $l_p$, 
then $C_n$ satisfies 
\begin{equation}\lb{HJ6}
\sqrt{1-\MO\l(\l(a/l_p\r)^{-2n+2}\r)} \lesssim C_n \leq 1.
\end{equation}

For $1<n\leq2$, we have $\MO\l(\l(a/l_p\r)^{-2n+2}\r) \ll 1$, which means $C_n \approx 1$ from \Ref{HJ6}, 
and we can conclude that the surface shrinks approximately at the speed of light as \Ref{eomR}. 
For $n=1$, on the other hand, we have $0<C_{n=1}\leq1$, which indicates that the surface can be timelike. 



\begin{thebibliography}{99}
\bibitem{foot:no_correct}
This is because a correction term, say $R^3$, contributes as 
$\MO(N^1)\f{g^{\mu\nu}(x)}{\sqrt{-g(x)}}\f{\d }{\d g^{\mu\nu}(x)} \int d^4x\sqrt{-g}R^3 
=\MO(N^1)(R(x)^3+9\Box R(x)^2) \sim N {\cal R} ^3 \sim \f{1}{N^2}$, 
which is negligible for a large $N$.
\bibitem{foot:a-dep}
In the limit $a\to\infty$, the exponentially large redshift becomes infinite (see \Ref{interior}),  
and the time inside is completely frozen, leading to the $a$-independence of $g^{(in)}_{\mu\nu}$ exactly. 
For a large but finite $a(\gg l_p)$, 
$A_0$ and $B_0$ in \Ref{AB0} may be slightly dependent on $a$ \ci{KY1}. 
However, such $a$-dependence is a subleading effect for $a\gg l_p$, 
and we focus on the leading one, $a$-independent part.
\bibitem{foot:AdS}
Putting $f(r)=\f{L^2}{z^2}$, we have $\p_r \log f(r) dr=-\f{2}{z}dz$.  
This leads to $h(r)dr^2=\f{L^2}{z^2}dz^2$ if \Ref{AdS} is satisfied.
\bibitem{foot:brick}
In the case of $N$ conformal scalar fields, if we assume $\eta=1$, 
\Ref{Ra} gives $R(a)=a+\f{N l_p^2}{360 \pi a}$. 
Here we have used the third equation of \Ref{AB_sol2} with $c_W=\f{N}{1920 \pi^2}$ \ci{BD,Duff}. 
This is the same as the location of the ``brickwall" \ci{brick,Frolov2}.
\bibitem{foot:u_derive}
From \Ref{interiort}, we can derive \Ref{interior2} as follows:   
\begin{align}
ds^2=&-\f{2\s}{r^2}e^{-\f{R(a(t))^2-r^2}{2\s\eta}}dt^2+\f{r^2}{2\s}dr^2+r^2d\Omega^2\nn\\
&=-\f{2\s}{r^2}e^{\f{r^2}{2\s\eta}} \l(e^{-\f{R(a(t))^2}{4\s\eta}}dt + \f{r^2}{2\s}e^{-\f{r^2}{4\s\eta}}dr\r)\times\nn\\
&~~~\l(e^{-\f{R(a(t))^2}{4\s\eta}}dt - \f{r^2}{2\s}e^{-\f{r^2}{4\s\eta}}dr\r)+r^2d\Omega^2\nn\\
\lb{interior2}
&=-\f{2\s}{r^2}e^{-\f{R(a(u))^2-r^2}{2\s\eta}}du^2-2e^{-\f{R(a(u))^2-r^2}{4\s\eta}}dudr+r^2d\Omega^2.\nn
\end{align}
As \Ref{dat} comes from \Ref{eomR}, which means that the surface is null, 
\Ref{da} is obtained by setting $ds^2|_{r=R(a(u))}=0$ in \Ref{interior2} 
and using $R(a(u))\approx a(u)$.
\bibitem{foot:sigma2}
$\s_s$, \Ref{sigma_s}, doesn't agree with $\s$ for scalar fields in \Ref{AB_sol2}. 
This is natural because $\s_s$ is obtained by considering s-waves in the eikonal approximation 
while $\s$ comes from all the modes with arbitrary angular momentum. 
See \ci{KY3,KY4} for the self-consistency of $\s$.
\bibitem{foot:differ}
This Hawking-like radiation occurs by a different mechanism from the usual one, 
although the magnitudes are the same \ci{foot:sigma1}. 
See \ci{KY4} for details. 
\end{thebibliography}
\end{document}